\documentclass[sigconf]{acmart}   % review
\usepackage[ruled,vlined]{algorithm2e}
\usepackage{amsmath}
\DeclareMathOperator*{\argmin}{arg\,min}
\DeclareMathAlphabet{\altmathcal}{OMS}{cmsy}{m}{n}
\usepackage{bm}
\usepackage{amsfonts}
\usepackage{enumitem}

\usepackage{makecell}

\newcommand{\dataset}{\textsc{GUIDE}}
\newcommand{\methodshort}{\textsc{TITAN}}
\newcommand{\method}{\textsc{Threat Intelligence Tracking via Adaptive Networks}}

\newcommand{\hide}[1]{}

\AtBeginDocument{%
  }

\setcopyright{acmlicensed}
\copyrightyear{2024}
\acmYear{2024}
\acmDOI{XXXXXXX.XXXXXXX}

\acmISBN{978-1-4503-XXXX-X/18/06}

\begin{document}

%%
%% The "title" command has an optional parameter,
%% allowing the author to define a "short title" to be used in page headers.
\title{Web Scale Graph Mining for Cyber Threat Intelligence}

%%
%% The "author" command and its associated commands are used to define
%% the authors and their affiliations.
%% Of note is the shared affiliation of the first two authors, and the
%% "authornote" and "authornotemark" commands
%% used to denote shared contribution to the research.
\author{Scott Freitas}
\email{scottfreitas@microsoft.com}
% \orcid{}
\affiliation{%
  \institution{Microsoft Security Research}
  \city{Redmond}
  \state{WA}
  \country{USA}
}

\author{Amir Gharib}
\email{agharib@microsoft.com}
\affiliation{%
  \institution{Microsoft Security Research}
  \city{Toronto}
  \state{}
  \country{Canada}
}

%%
%% By default, the full list of authors will be used in the page
%% headers. Often, this list is too long, and will overlap
%% other information printed in the page headers. This command allows
%% the author to define a more concise list
%% of authors' names for this purpose.
\renewcommand{\shortauthors}{Freitas \& Gharib}

\begin{abstract}
Defending against today’s increasingly sophisticated and large-scale cyberattacks demands accurate, real-time threat intelligence. 
Traditional approaches struggle to scale, integrate diverse telemetry, and adapt to a constantly evolving security landscape. 
We introduce \method{} (\methodshort{}), an industry-scale graph mining framework that generates cyber threat intelligence at unprecedented speed and scale. 
\methodshort{} introduces a suite of innovations specifically designed to address the complexities of the modern security landscape, including:
(1) a dynamic threat intelligence graph that maps the intricate relationships between millions of entities, incidents, and organizations;
(2) real-time update mechanisms that automatically decay and prune outdated intel;
(3) integration of security domain knowledge to bootstrap initial reputation scores;
and (4) reputation propagation algorithms that uncover hidden threat actor infrastructure. 
Integrated into Microsoft Unified Security Operations Platform (USOP), which is deployed across hundreds of thousands of organizations worldwide, \methodshort's threat intelligence powers key detection and disruption capabilities.
With an impressive average macro-F1 score of $0.89$ and a precision-recall AUC of $0.94$, \methodshort{} identifies millions of high-risk entities each week, enabling a 6x increase in non-file threat intelligence. 
Since its deployment, \methodshort{} has increased the product's incident disruption rate by a remarkable $21\%$, while reducing the time to disrupt by a factor of 1.9x, and maintaining $99\%$ precision, as confirmed by customer feedback and thorough manual evaluation by security experts---ultimately saving customers from costly security breaches.
\end{abstract}

%%
%% The code below is generated by the tool at http://dl.acm.org/ccs.cfm.
%% Please copy and paste the code instead of the example below.
%%


%%
%% Keywords. The author(s) should pick words that accurately describe
%% the work being presented. Separate the keywords with commas.

% \received{8 August 2024}
% \received[revised]{12 March 2009}
% \received[accepted]{5 June 2009}

%%
%% This command processes the author and affiliation and title
%% information and builds the first part of the formatted document.
\maketitle

\section{Introduction}
In today’s cybersecurity landscape, threat actors continuously evolve their techniques to infiltrate networks by leveraging a vast array of interconnected infrastructure. 
This has created an urgent demand for high-quality, real-time threat intelligence (TI).
However, traditional TI approaches often struggle to scale, relying on manual investigation, signature matching, static analysis, and behavioral monitoring~\cite{palo2024intelligence,google2024intelligence,virus2024intelligence}.
These methods are further hindered by their siloed nature, lacking broader context across the entire enterprise security landscape, resulting in a fragmented view of threat actor infrastructure~\cite{chau2011polonium,tamersoy2014guilt}.

Unified security operation platforms platforms, such as Microsoft USOP, are uniquely positioned to break down these silos by acting as the centralized security hub.
These platforms aim to enhance efficiency and effectiveness by correlating alerts across first and third party security products, such as endpoint, email, and identity, into cohesive security incidents ~\cite{einav2023introducing,freitas2024graphweaver}.
With \methodshort{}, we advance Microsoft USOP threat intelligence capabilities by introducing a real-time, dynamic TI graph that captures the complex relationships between millions of entities, incidents, and organizations, providing a unified view of threat activity.
By infusing this graph with security domain knowledge and leveraging a guilt-by-association framework~\cite{koutra2011unifying}, we propagate reputation scores to unknown entities, enabling early detection and disruption (i.e., pre-damage mitigation) of threat actor infrastructure.

\begin{figure*}[t!]
    \centering
    \includegraphics[width=\textwidth]{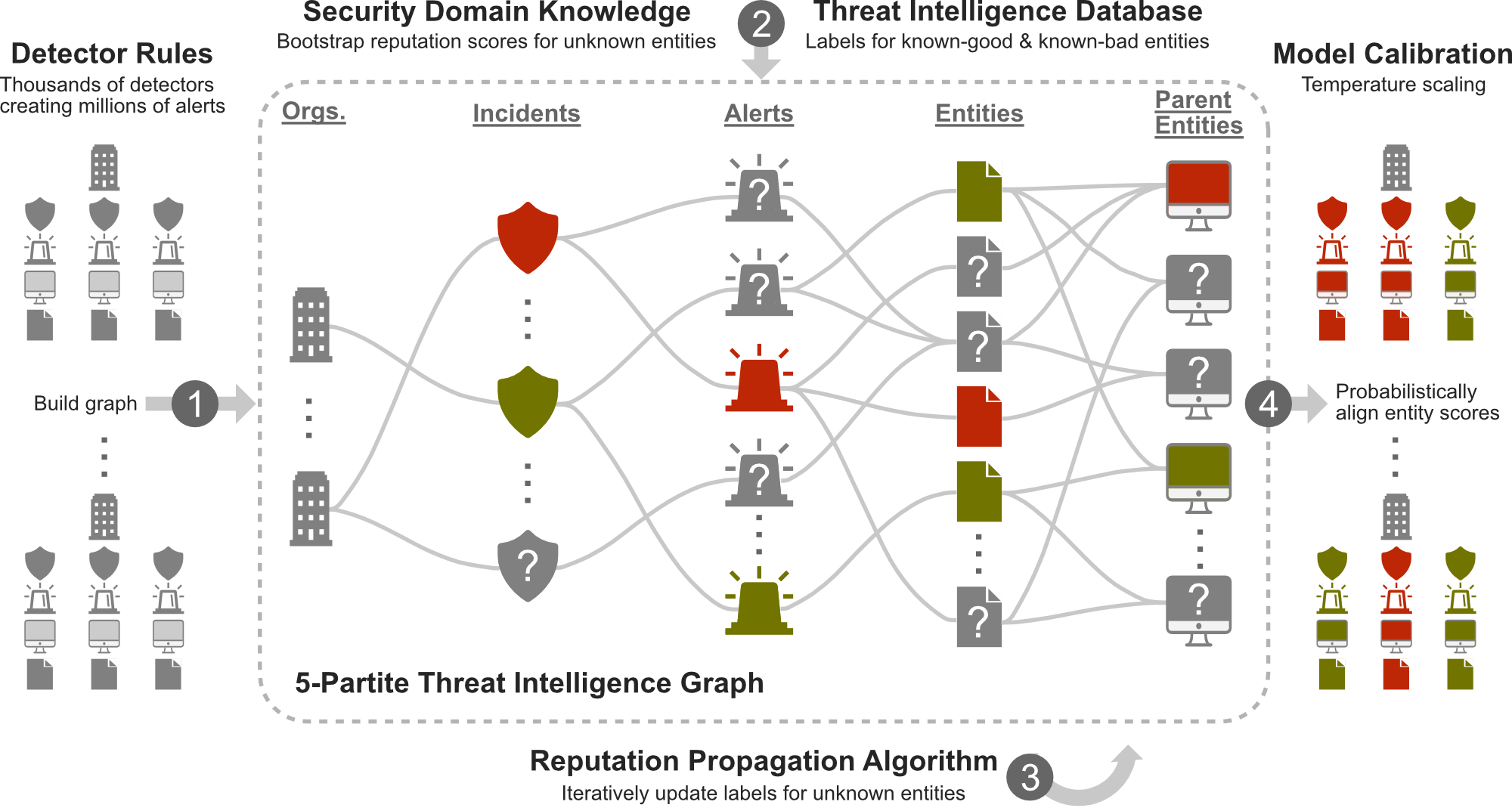}
    \caption{Overview of the \methodshort{} architecture: an industry-scale graph mining framework that generates real-time TI by propagating reputation scores across millions of interconnected entities, incidents, and organizations. 
    Built on a time evolving 5-partite graph, the system operates through four key components: 
    (1) dynamic graph construction and updates,
    (2) integration of known TI and security domain knowledge to bootstrap reputation scores for unknown entities;
    (3) reputation propagation to iteratively update risk scores;
    and (4) model calibration to probabilistically align scores for use by security analysts.
    }
    \label{fig:crown}
\end{figure*}

\vspace{1mm}\noindent
\textbf{Threat intelligence at scale.} Generating scalable and accurate threat intelligence presents multiple unique and exciting challenges:

\begin{enumerate}[ topsep=4pt, leftmargin=*, itemsep=3pt]
    \item \textbf{Evolving threat environment.}
    Adversaries continually evolve their tactics and infrastructure, creating a rapidly shifting threat landscape. 
    Generating up-to-date intelligence while identifying and pruning stale data is a substantial challenge.
        
    \item \textbf{Complex security landscape.} 
    The vast array of commercial security products, each with thousands of custom and built-in detection rules, creates an intricate and fragmented enterprise environment. 
    Integrating diverse security telemetry into a unified TI framework requires careful application of domain knowledge to ensure accurate and meaningful insights.

    \item \textbf{Scalable and robust architecture.}
    Modern security systems generate enormous volumes of alerts across interconnected domains such as network, cloud, endpoint, and email. 
    Scaling to analyze millions of entities and terabytes of data in real time demands a robust, low-latency, and efficient architecture. 
\end{enumerate}

The emergence of USOP as a relatively new industry underscores the timeliness of these challenges and positions scalable threat intelligence as a pivotal frontier in cybersecurity. 
Innovative solutions that drive real-time TI generation will be essential to safeguarding organizations against continuously evolving threats.

\subsection{Contributions}
We introduce \methodshort{} (Figure~\ref{fig:crown}), a novel framework designed to address the challenges of generating high-fidelity threat intelligence at scale and in real time. 
Our framework makes significant contributions in the following areas:

\begin{itemize}[topsep=2mm, itemsep=0mm, parsep=1mm, leftmargin=*]
    \item \textbf{\methodshort{} architecture.} 
    \methodshort{} transforms the cybersecurity industry's approach to threat intelligence by introducing advanced methods for real-time, large-scale TI generation. 
    Key innovations include: (1) dynamic k-partite graph that captures complex relationships between entities, incidents, and organizations; 
    (2)the integration of security domain knowledge to bootstrap initial reputation scores; 
    (3) reputation propagation algorithms to uncover hidden threat actor infrastructure; 
    and (4) model calibration to probabilistic align reputation scores. 
    We also disclose key architectural design elements and operational processes, setting a precedent as the first USOP cybersecurity company to openly discuss advanced TI capabilities in such comprehensive detail.
    
    \item \textbf{Extensive Evaluation.}
    We conduct a comprehensive evaluation of \methodshort{}'s performance across three key pillars: internal assessments, collaborations with security experts, and customer feedback. 
    In internal testing on hundreds of thousands of held out entities, \methodshort{} achieves an impressive average cross-region macro-F1 score of $0.89$ and a precision-recall AUC of $0.94$. 
    
    \item \textbf{Impact to Microsoft Customers and Beyond.} 
    \methodshort{} is integrated into Microsoft USOP, a market leader~\cite{mellen2024forrester}, deployed across hundreds of thousands of organizations worldwide.
    Each week, \methodshort{} identifies millions of high-risk entities, enabling a 6x increase in non-file threat intelligence.
    This research has transformed the product’s approach to detection and disruption, increasing the overall incident disruption rate by $21\%$ while reducing the time to disrupt by a factor of 1.9x---saving customers from costly breaches.
    Collaboration with Microsoft security research experts and feedback from customers further validates the effectiveness of our TI, demonstrating $99\%$ precision in attack disruption scenarios.
\end{itemize}

\begin{table}[t]
\centering
\begin{tabular}{p{0.085\textwidth} p{0.35\textwidth}}

\textbf{Term} & \textbf{Definition} \\
\cmidrule(r){1-1} \cmidrule(lr){2-2}
Alert & Potential security threat that was detected  \\ \addlinespace
Detector & A security rule or ML model that generates alerts \\ \addlinespace
Entity & File, IP, etc. evidence associated with an alert \\ \addlinespace
Correlation & A link between two alerts based on a shared entity \\ \addlinespace
Incident & Related alerts that are correlated together \\ \addlinespace
Organization & Company containing a USOP product \\ \addlinespace
Disrupted & Early threat mitigation (e.g., disable user) \\ \addlinespace
Reputation & Likelihood of an entity being malicious or benign \\ \addlinespace
USOP & Unified Security Operations Platforms (USOP) are used to protect organizations across the entire 1st and 3rd party enterprise landscape 
\\ \addlinespace
\bottomrule
\end{tabular}
\caption{Terminology and definitions.}
\vspace{-5mm}
\label{table:terminology}
\end{table}
\section{Related Work}
We review key research related to \methodshort{}, including: threat intelligence platforms in Sec.~\ref{subsec:platform-related}, security knowledge graphs in Sec.~\ref{subsec:knowledge-related}, and graph-based reputation propagation mechanisms in Sec.~\ref{subsec:guilt-related}. 
We also highlight how \methodshort{} differentiates itself within the TI landscape.
To enhance readability, Table~\ref{table:terminology} details the terminology used in this paper.

\subsection{Threat Intelligence Platforms}\label{subsec:platform-related}
Threat intelligence platforms play a key role in collecting, analyzing, and distributing threat data to help organizations stay ahead of cyber threats. 
Platforms such as \cite{ibm2024intelligence,palo2024intelligence,google2024intelligence,berg2024mdti} aggregate open-source intelligence (OSINT), proprietary product telemetry, and human analysis to track threat actor infrastructure, vulnerabilities, and indicators of compromise. 
Alternatively, \cite{crowdstrike2024intelligence} focuses on dark web monitoring and real-time adversary tracking, offering insights into cyber attackers' tactics, techniques, and procedures. 
Others like \cite{virus2024intelligence} provide actionable intelligence by analyzing files, URLs, and other artifacts for malicious activity.
Additionally, \cite{recorded2024intelligence} integrates machine learning with human expertise to build an intelligence graph that monitors threats across OSINT and the dark web.
While these platforms and many others provide essential services, they often rely heavily on human investigation, and when machine learning is applied, the underlying mechanisms are not disclosed. 
In contrast, \methodshort{} advances the Microsoft Unified Security Operations Platform by providing the first transparent view of an industry-scale graph mining system that generates real-time TI, setting a new standard for openness in the field.

\subsection{Security Knowledge Graphs (SKGs)}\label{subsec:knowledge-related}
SKGs aggregate and analyze threat intelligence by representing the relationships between entities such as malware, software vulnerabilities, and attack patterns~\cite{liu2022review}. 
These entities are drawn from a variety of structured and unstructured sources, including blogs, security bulletins, CVEs, malware reports, as well as CTI and APT reports~\cite{hu2024llm,rastogi2020malont,gascon2017mining,sarhan2021open,gao2021enabling}. 
Extensive research has focused on automating the extraction of these entities and relationships using techniques such as natural language processing and machine learning to create and analyze structured knowledge graphs~\cite{huangctikg,li2022attackg,piplai2020creating,gascon2017mining,rastogi2021information,zhao2020cyber,sarhan2021open}. 
Once constructed, SKGs are widely applied in areas across vulnerability analysis~\cite{noel2016cygraph,wu2017network}, threat discovery~\cite{narayanan2018early,gao2021enabling,daluwatta2022cgraph,gascon2017mining,modi2016towards}, decision-making~\cite{qi2023cybersecurity,mittal2019cyber,chen2021automatic}, and attack attribution~\cite{ren2022cskg4apt,christian2021ontology,duan2024heterogeneous,dutta2021malware,zhu2018cyber}.
While both SKGs and TITAN use graph-based structures, SKGs focus on converting \textit{already known information} into structured graphs for long-term knowledge storage and querying. 
In contrast, TITAN analyzes structured event and alert-level telemetry from security products and employs guilt-by-association models to dynamically propagate reputation scores and \textit{generate new threat intelligence} in real-time.

\subsection{Guilt By Association}\label{subsec:guilt-related}
Graph-based reputation propagation has played a key role in enhancing threat detection within Endpoint Detection and Response systems. 
By leveraging file-based relationship graphs alongside reputation propagation, these methods effectively identify malicious software~\cite{hu2014asset, najafi2024you, chen2015analyzing, chen2015intelligent, ni2015file, karampatziakis2013using, ye2011combining}. 
Industry solutions such as Mastino~\cite{rahbarinia2016real}, Polonium~\cite{chau2011polonium}, and AESOP~\cite{tamersoy2014guilt} have all implemented variations of graph-based guilt-by-association techniques to detect malware.
In addition, graph propagation techniques have been applied to identify malicious domains, IP addresses, and malware distribution infrastructure~\cite{najafi2018guilt, saha2014detecting, oprea2015detection, manadhata2014detecting, kim2022phishing, khalil2018domain, invernizzi2014nazca, najafi2019malrank, huang2015large}.
\methodshort{} builds on this foundational work by introducing a real-time, dynamic threat intelligence graph that captures the complex relationships across the entire enterprise landscape.

\section{\methodshort{} Graph Architecture}\label{sec:graph-overview}
\methodshort's graph architecture and design choices were carefully selected in close collaboration with security researchers, leveraging domain expertise to fine-tune how relationships, reputation scores, and entity relevance evolve over time. 
This section is structured around four components: 
formulation of the multi-partite graph 
in Sec~\ref{subsec:graph-formulation};
initialization of node reputation scores in Sec~\ref{subsec:graph-node-scores}; 
integration of security domain knowledge 
in Sec~\ref{subsec:graph-edge-weights}; and
dynamic batch updates and pruning of outdated telemetry in Sec~\ref{subsec:graph-dynamic-updates}.

\begin{table*}[h!]
\centering
\setlength{\tabcolsep}{6pt}
\begin{tabular}{lllllrrrr}
\toprule
\textbf{\makecell[l]{Graph\\Layers}} & \textbf{\makecell[l]{Source\\Node}} & \textbf{\makecell[l]{Target\\Node}} & \textbf{\makecell[l]{Target\\Description}} & \textbf{\makecell[l]{Decay\\Function}} & \textbf{\makecell[r]{Initial\\Weight}} & \textbf{\makecell[r]{Decay\\Rate (h)}} & \textbf{\makecell[r]{Max Alive\\Time (h)}}  & \textbf{\makecell[r]{Edge\\Dist. (\%)}} \\ 
\midrule

1 $\rightarrow$ 2 & OrgId & IncidentId & Org-level incident identifier & Constant & 0.1 & - & 168 & 2.5 \\ 

2 $\rightarrow$ 3 & IncidentId & AlertId & Unique alert identifier & Constant & 1 & - & 168 & 6.6 \\ 

3 $\rightarrow$ 4 & AlertId & SHA1 & Cryptographic file hash & Constant & 1 & - & 168 & .47 \\ 

3 $\rightarrow$ 4 & AlertId & CampaignId & Email campaign identifier  & Constant & 1 & - & 120 & .05\\ 

3 $\rightarrow$ 4 & AlertId & SessionId & Cloud session identifier & Constant & 1 & - & 168 & .02 \\ 

3 $\rightarrow$ 4 & AlertId & EmailId & Email message identifier & Constant & 1 & - & 24 & 12 \\ 

3 $\rightarrow$ 4 & AlertId & AppId & Identifier for cloud app & Linear  & 1 & - & 48 & $1\mathrm{e}{-3}$ \\ 

3 $\rightarrow$ 4 & AlertId & URL & Website URL link & Linear & 1 & - & 12 & .32 \\ 

3 $\rightarrow$ 4 & AlertId & IpAddress & Public IP address & Linear  & 1 & - & 12 & 5.3 \\ 

3 $\rightarrow$ 4 & AlertId & DeviceName & Identifier for device & Exponential & 1 & 0.19 & 12 & .31 \\ 

3 $\rightarrow$ 4 & AlertId & ResourceId & Cloud resource identifier & Exponential & 1 & 0.24 & 12 & $1\mathrm{e}{-4}$  \\ 

3 $\rightarrow$ 4 & AlertId & RegistryKey & OS registry key & Exponential & 1 & 0.24 & 12 & .02 \\ 

3 $\rightarrow$ 4 & AlertId & RegistryVal & Data stored in key & Exponential & 1 & 0.24 & 12 & .02 \\ 

4 $\rightarrow$ 5 & SHA1 & FileDir &  File directory path & Exponential & 1 & 0.24 & 24 & .25 \\ 

4 $\rightarrow$ 5 & EmailId & EmailAddress & Email sender address & Linear & 1 & - & 12 & 1.0 \\ 

4 $\rightarrow$ 5 & EmailId & EmailCluster & Email cluster identifier & Linear & 0.5 & - & 6 & 19 \\ 

4 $\rightarrow$ 5 & URL & URLDomain & Domain name of URL & Linear & 0.5 & - & 6 & .05 \\ 

4 $\rightarrow$ 5 & IpAddress & IpRange & Public IPs in subnet /24 & Linear & 0.5 & - & 12 & .32 \\ 

\bottomrule
\end{tabular}
\caption{Graph structure representing the hierarchical relationships between nodes. 
For example, $1\rightarrow2$ signifies a connection between an organization (layer 1) and an incident (layer 2). 
Each edge is characterized by its decay function, initial weight, decay rate, and maximum alive time, which govern how relationships between nodes evolve over time. 
The edge distribution reflects the prevalence of each edge type averaged across 782 cross-region graph snapshots.
}
\label{tab:edge-types}
\vspace{-4mm}
\end{table*}

\subsection{Multi-Partite Graph Formulation}\label{subsec:graph-formulation}
The graph is designed as a dynamic, time-evolving, undirected, weighted k-partite graph, comprising five hierarchical node layers that represent key components of the enterprise security landscape: (1) organizations, (2) incidents, (3) alerts, (4) entities, and (5) parent entities. 
Figure~\ref{fig:crown} visually depicts this graph structure, while Table~\ref{tab:edge-types} details the hierarchical relationships between node layers.

\smallskip\noindent
\textbf{K-partite topology.}
The graph topology in Table~\ref{tab:edge-types} captures the natural hierarchy in security telemetry, from organizations and incidents down to individual entities like files, IPs, and emails. 
The graph layers column illustrates these connections, while the edge distribution column shows average edge prevalence across 782 cross-region graph snapshots.
Email related edges are the most common, driven by the high volume of phishing attacks.

\smallskip\noindent
\textbf{Undirected graph.}
An undirected graph topology provides flexibility in representing relationships between nodes, where interactions are not inherently one-way. 
In cybersecurity, many entities such as IP addresses, files, and alerts can influence and be influenced in a reciprocal manner.
Additionally, undirected graphs help avoid the challenges associated with asymmetric information flow found in directed graphs, such as cycles and spider traps.

\smallskip\noindent
\textbf{Nodes and edges.}
Nodes and edges in the graph represent the relationships between entities, alerts, incidents, and organizations, with telemetry sourced from customer-developed detection rules, Microsoft security products, and third-party security solutions.
We profile 16 infrastructure entities that can be compromised or exploited by threat actors, as outlined in Table~\ref{tab:edge-types}. 
\methodshort's flexible graph architecture accommodates new entity types with minimal modifications, allowing for the integration of additional nodes and edges as the threat landscape evolves.

\subsection{Initializing Reputation Scores}\label{subsec:graph-node-scores}
Each node in the graph is initialized with a reputation score between $0$ and $1$, representing its likelihood of being benign ($0$), malicious ($1$), or unknown ($0.5$). 
Alert nodes are initially assigned a default score of $0.5$ to prevent overly aggressive labeling, given their central role in propagation.
Alerts that are automatically disrupted~\cite{pawel2023xdr} are also assigned a score of $1$.
Incident nodes inherit the highest score from their alerts, while organizational nodes start at $0.5$ due to their downstream influence. 
Entity scores are derived from expert vetting, detection rules, and bootstrapped cross-organizational reputation scoring.
Nodes with scores $\geq 0.9$ are classified as malicious, $\leq 0.1$ as benign, and others as unknown. 
As shown in Table~\ref{table:combined-results}, on average, 3\% of entity nodes are malicious, 39\% are unknown, and 58\% are benign, with a large proportion of benign initializations coming from third party telemetry.

\smallskip\noindent
\textbf{Bootstrapped reputation scoring.}
Threat actors often reuse infrastructure across attack campaigns, allowing us to aggregate intelligence from multiple organizations 
to bootstrap high-confidence reputation scores based on factors such as:

\begin{itemize}[topsep=4pt, leftmargin=*, itemsep=3pt]
    \item Number of organizations identifying the entity as true positive
    \item Number of security products flagging the entity
    \item Number of unique detection rules linked to the entity
    \item Number of alerts associated with the entity
    \item Number of organizations observing the entity
    \item Number of true positive graded alerts
    \item Number of false positive or benign positive graded alerts
\end{itemize}

These metrics are aggregated to produce a composite reputation score for each entity ranging from 0 to 1.
While we are not able to disclose the exact weighting of each component, the outlined factors offer key insight into the underlying scoring mechanism.

\subsection{Infusing Security Domain Knowledge}\label{subsec:graph-edge-weights}
In the dynamic TI graph, each edge is assigned a weight representing the strength of the relationship. 
To maintain the relevance of these relationships, we introduce edge weight decay functions (constant, linear, or exponential) that reduce an edge's weight based on the time elapsed since its creation.
Table~\ref{tab:edge-types} outlines the decay functions, initial weights, decay rates, and maximum alive times for each edge type in the graph. 
The initial weight reflects the importance of a relationship at the time of creation, the decay rate dictates how quickly its relevance diminishes, and the maximum alive time defines the lifespan of an entity before mandatory pruning. 
While the decay function parameters are based on realistic experimental settings, they are further refined in production, with the final values remaining confidential.

\subsection{Dynamic Graph Updates}\label{subsec:graph-dynamic-updates}
To maintain up-to-date threat intelligence while managing memory and computational efficiency, we incrementally update the graph by incorporating new telemetry at 60-minute intervals.
As the graph evolves, we prevent the accumulation of outdated information through a pruning process that removes edges with weights below a threshold ($0.01$), isolated nodes, and nodes exceeding their maximum lifespan (see Table~\ref{tab:edge-types}). 
This ensures the graph remains focused on relevant connections, supporting scalability across thousands of organizations and millions of nodes and edges.

\section{Uncovering Threat Actor Infrastructure}\label{sec:guilt-by-association}
Building on the graph structure outlined in Section~\ref{sec:graph-overview}, \methodshort{} employs a ``guilt by association'' framework, where unknown entities inherit reputation scores through connections to known malicious or benign entities, enabling us to uncover hidden threat actor infrastructure. 
Section~\ref{subsec:comparing-algorithms} compares reputation propagation methods, Section~\ref{subsec:label-propagation} details the propagation algorithm applied to the TI graph, and Section~\ref{subsec:calibration-scores} discusses model calibration to ensure probabilistic score interpretation. Algorithm~\ref{alg:label-propagation} outlines the entire process.

\subsection{Reputation Propagation Mechanisms.}\label{subsec:comparing-algorithms}
We considered four potential reputation propagation mechanisms,
each with their own advantages:

\begin{itemize}[topsep=4pt, leftmargin=*, itemsep=3pt]
    \item \textbf{Label propagation} (LP) offers excellent convergence in terms of speed and predictability, and is the simplest of the methods~\cite{zhuѓ2002learning}.

    \item \textbf{Label spreading} (LS) is effective in noisy environments when initial labels are unreliable, but requires parameter tuning~\cite{zhou2003learning}.

    \item \textbf{Loopy belief propagation} (LBP) excels at handling complex graph structures, but does not scale well to large graphs~\cite{murphy2013loopy}.

    \item \textbf{Graph neural networks} (GNNs) offer exceptional node classification capabilities~\cite{duan2022comprehensive,wu2020comprehensive}. However, we do not adopt them due to incompatibility with our  CPU-based PySpark infrastructure.
\end{itemize}

Upon evaluating each method in our production PySpark environment, we selected LP as the primary propagation mechanism. 
LP demonstrated superior scalability and convergence on our large-scale TI graph compared to LBP (over $5x$ faster), and matched LS in performance with line search parameter tuning.
As our proposed architecture is agnostic to the specific propagation mechanism, GNNs represent a promising future direction to explore.

\subsection{Label Propagation for Reputation Scoring}\label{subsec:label-propagation}
Label propagation is a semi-supervised technique that iteratively assigns labels to nodes based on the labels of their neighbors, refining the graph label distribution until convergence. 
Leveraging the TI graph created in Section~\ref{sec:graph-overview}, we convert it into a sparse weighted adjacency matrix $\bm{A}$ with an initial label matrix $\bm{Y}$ containing reputation scores for each node in the graph.
The implementation of Algorithm~\ref{alg:label-propagation} is discussed below. 
We adopt standard notation conventions, with capital bold letters for matrices (e.g., $\bm{A}$), and lowercase bold letters for vectors (e.g., $\bm{a}$).

\smallskip\noindent
\textbf{Initialization.}
We begin with an initial label matrix $\bm{Y}^{(0)}$, where each row corresponds to a node, and the two columns represent the probabilities of the node being benign (0) or malicious (1).
A degree matrix $\bm{D}$ is computed, which contains the sum of the weights of edges connected to each node.
The adjacency matrix $\bm{A}$ is normalized using $\hat{\bm{A}} = \bm{D}^{-1} \bm{A}$, which helps to evenly distribute influence across nodes.
A mask $\bm{M}$ is created to identify nodes with high-confidence labels (i.e., those with initial scores $\geq 0.9$  or $\leq 0.1$), ensuring ground-truth labels remain unchanged across iterations.

\smallskip\noindent
\textbf{Label propagation.}
In each iteration, the label matrix $\bm{Y}^{(t)}$ is updated by multiplying the normalized adjacency matrix $\hat{\bm{A}}$ with the previous iteration’s label matrix $\bm{Y}^{(t-1)}$.
This update effectively averages the labels of a node’s neighbors, spreading the label information across the graph.

\smallskip\noindent
\textbf{Applying the mask.}
After updating the labels, the mask $\bm{M}$ is applied to ensure high-confidence labels remain fixed. 
This is done by replacing the corresponding entries in $\bm{Y}^{(t)}$ with the original values from $\bm{Y}^{(0)}$ where the mask indicates they should not change.

\smallskip\noindent
\textbf{Row normalization.}
To maintain a valid probability distribution, each row of the label matrix $\bm{Y}^{(t)}$ is normalized so that the sum of its elements equals $1$.

\smallskip\noindent
\textbf{Convergence check.}
The algorithm checks for convergence by computing the Frobenius norm, repeating for up to $k = 100$ iterations or until it reaches the specified tolerance $\epsilon = 0.001$.

\begin{algorithm}[t]
\caption{Guilt-by-Association Reputation Modeling}
\label{alg:label-propagation}
\KwIn{Adjacency matrix $\bm{A}$, initial label matrix $\bm{Y}^{(0)}$, max iters. \( k \), tolerance \( \epsilon \), min \& max temps $T_{min}$, $T_{max}$}
\KwOut{Calibrated reputation scores \( \bm{Y}_{T^*} \)}
\BlankLine

\SetKwBlock{Initialization}{Initialization}{end}
\SetKwBlock{Propagation}{Label Propagation}{end}
\SetKwBlock{Temperature}{Temperature Scaling}{end}

\Initialization{
    Compute degree matrix \( \bm{D} \) with entries \( \bm{D}_{ii} = \sum_{j} \bm{A}_{ij} \) \\
    Compute normalized adjacency matrix \( \bm{\hat{A}} = \bm{D}^{-1} \bm{A} \) \\
    Define mask \( \bm{M} \) where \( \bm{M}_{ij} = 1 \) if \( \bm{Y}^{(0)}_{ij} \geq 0.9 \) or \( \bm{Y}^{(0)}_{ij} \leq 0.1 \), otherwise \( \bm{M}_{ij} = 0 \) \\[0.5em]
}

\BlankLine

\Propagation{
    \For{\( t = 1 \) \textbf{to} \( k \)}{
        $\bm{Y}^{(t)} = \bm{\hat{A}} \bm{Y}^{(t-1)}$  \tcp*{Update labels}
        
        $\bm{Y}^{(t)} = \bm{M} \cdot \bm{Y}^{(0)} + (1 - \bm{M}) \cdot \bm{Y}^{(t)}$ \tcp*{Apply mask}

        $\bm{Y}^{(t)} \leftarrow \bm{D}^{-1} \bm{Y}^{(t)}$ \tcp*{Normalize rows}
        
        \If{\( \Delta = \| \bm{Y}^{(t)} - \bm{Y}^{(t-1)} \|_F < \epsilon \)}{
            \textbf{Break} \tcp*{Convergence} 
        }
    }
}

\Temperature{
    $\bm{Y}_T = \text{softmax}\left(\frac{\bm{Y^{(t)}}}{T}\right)$ \tcp*{Scale logits}

    \BlankLine

    $\text{NLL}(T) = -\frac{1}{N} \sum\limits_{n=1}^N \sum\limits_{i=1}^C \bm{Y}_{ni} \log\left( \bm{Y}_{T,ni} \right)$ 

    \BlankLine
    
    $T^* = \argmin_{T \in [T_{\text{min}}, T_{\text{max}}]} \text{NLL}(T)$ \tcp*{Minimize NLL}

    \BlankLine

    $\bm{Y}_{T^*} = \text{softmax}\left( \frac{\bm{Y}^{(t)}}{T^*} \right)$ \tcp*{Calibrate scores}

}

\end{algorithm}

\subsection{Calibrating Reputation Scores}\label{subsec:calibration-scores}
Graph-based propagation algorithms distribute reputation scores by leveraging relationships between nodes to infer scores for those with unknown values. 
The algorithm generates raw scores, or logits, which reflect the confidence in assigning labels to each node. 
However, these uncalibrated logits often result in inaccurate probabilistic interpretations~\cite{liu2022calibration,hsu2022makes}, a critical limitation when security analysts rely on this information to make high-stakes decisions.

\smallskip\noindent
\textbf{Temperature scaling.}
To address this issue, we apply temperature scaling~\cite{guo2017calibration}, a technique that ``softens'' logits by introducing a temperature parameter $T$, which adjusts the sharpness of the predicted probabilities. 
When $T > 1$, the logits are scaled down, producing softer and less extreme probabilities. 
Conversely, when $T < 1$, the logits are scaled up, making the probabilities more confident.
For each node $x$ in the graph, let $z_i(x)$ denote the logit for class $i$. 
To apply temperature scaling, we divide the logits by the temperature parameter $T$, which are then used to compute the calibrated probabilities via the softmax function.

\smallskip\noindent
\textbf{Minimize NLL.}
The optimal temperature $T$ is determined by minimizing the negative log-likelihood (NLL) on a validation set. 
Let $N$ be the number of samples, $C$ the number of classes, and $y_{n,i}(x)$ the true label indicator (1 for malicious, 0 for benign) for each sample $n$ and class $i$.
The NLL is minimized over the temperature range $T_{\text{min}} = 0.1$ to $T_{\text{max}} = 10$ using the L-BFGS-B method~\cite{byrd1995limited}.
This process results in an optimal temperature $T^*$ that produces the best-calibrated probabilities based on the validation set.

\smallskip\noindent
\textbf{Score calibration.}
Once the optimal temperature $T^*$ is found, it is used to scale the predicted reputation scores, yielding scores that are probabilistically interpretable by security analysts.

\section{Experiments}
We present \methodshort's experimental framework and findings. 
Section~\ref{subsec:experimental-setup} describes the setup for evaluating \methodshort's performance. 
Section~\ref{subsec:graph-analysis} explores how temporal factors influence the graph. 
Section~\ref{subsec:reputation-scoring} analyzes \methodshort's effectiveness in uncovering threat actor infrastructure.
Finally, Section~\ref{subsec:model-calibration} assesses the ability to probabilistically align reputation scores.

\subsection{Experimental Setup}\label{subsec:experimental-setup}
To comply with privacy regulations, \methodshort{} is uniformly replicated across geographic regions. 
We evaluate performance across a sample of $12$ regions, each with ground-truth reputation scores randomly split into $70\%$ for reputation propagation, $10\%$ for calibration, and $20\%$ for evaluation. 
Ground truth is derived from signature-based matching, expert validation, customer feedback, detectors with a $99\%$ signal-to-noise ratio, and bootstrapped scores as outlined in Section~\ref{subsec:graph-node-scores}. 
Performance is measured using macro-F1, along with precision and recall, as is standard for imbalanced datasets~\cite{freitas2022malnet,freitas2021large,duggal2021har,duggal2020rest}.

\medskip\noindent
\textbf{GUIDE dataset.}
Introduced in our prior work, the \dataset{} dataset \cite{freitas2024ai} comprises over 13 million data points across 33 entity types, including 1.6 million alerts and 1 million incidents labeled with customer-provided triage and remediation responses collected over two weeks. 
Sourced from 6.1k organizations and covering 9.1k unique detector types across various security products, it is publicly available on \href{https://www.kaggle.com/datasets/Microsoft/microsoft-security-incident-prediction}{Kaggle}.
\dataset{} is sourced from telemetry in Region 1, the focus of our analysis, enabling \methodshort{} to provide a foundational baseline for future advancements in threat intelligence research.

\subsection{Temporal Graph Dynamics}\label{subsec:graph-analysis}
\begin{figure}[t]
    \centering
    \includegraphics[width=1\linewidth]{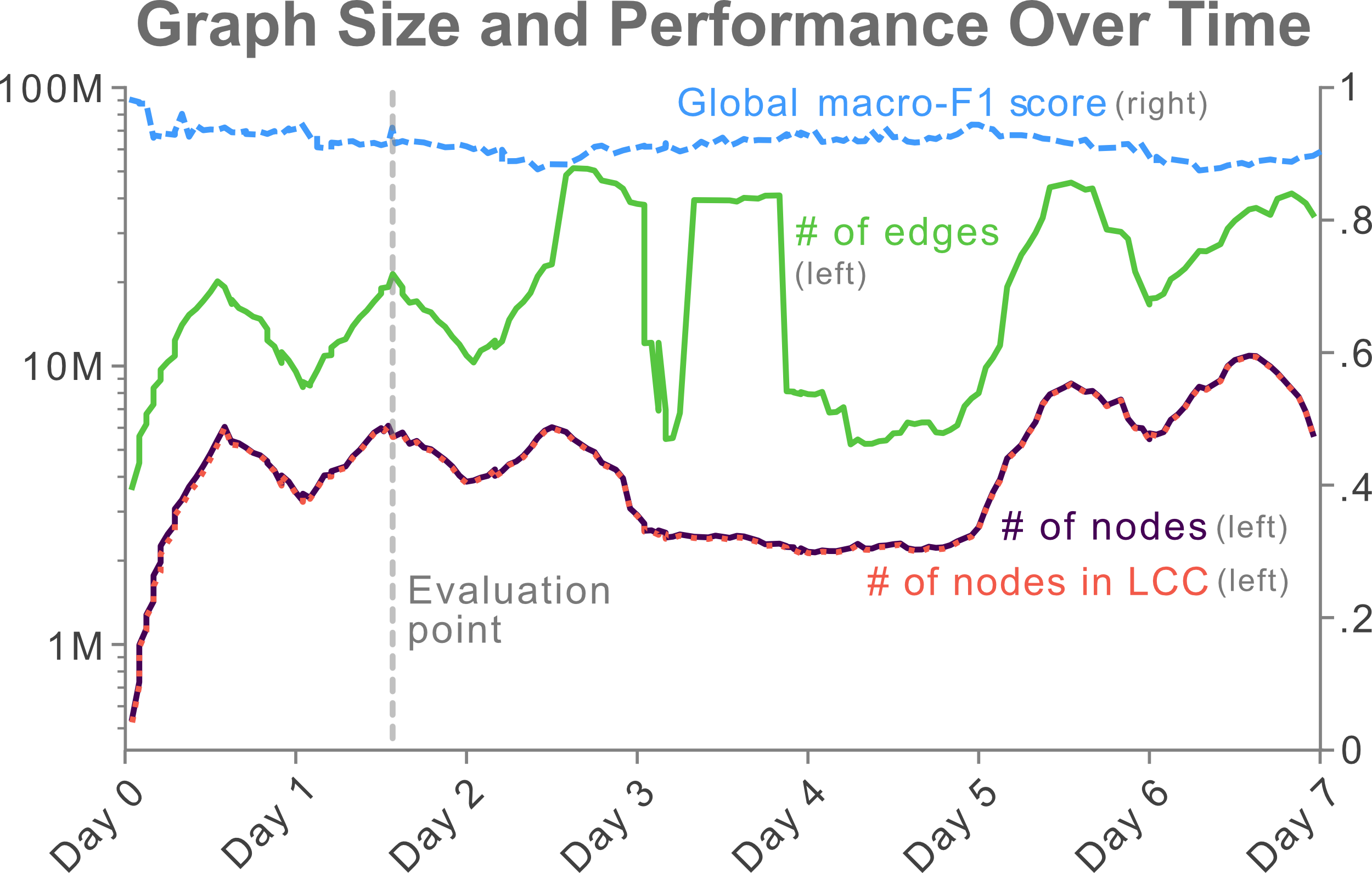}
    \caption{Graph size and performance over a $7$ day period. 
    The left y-axis plots the number of nodes, nodes in the largest connected component (LCC), and edges. 
    The right y-axis captures the average macro-F1 score across all entities (global).}
    \label{fig:graph-size}
    \vspace{-2mm}
\end{figure}

We examine the temporal dynamics of the TI graph to understand how it scales and adapts to an evolving threat landscape. Figure~\ref{fig:graph-size} illustrates the evolution of Region 1 over a seven-day period, tracking nodes, edges, and the largest connected component (LCC). 
Node and edge volume fluctuates due to telemetry updates, time of day, and the pruning of outdated data. 
Despite these changes, the LCC remains proportionally large, preserving a single giant connected component. 
The global macro-F1 score averages $0.92$ across runs, demonstrating that structural adjustments do not impact the algorithm’s ability to propagate information.
Our subsequent analysis centers on the ``evaluation point'' in Figure~\ref{fig:graph-size}, providing a representative snapshot of the temporal graph dynamics.

\begin{figure}[b]
    \centering
    \includegraphics[width=\linewidth]{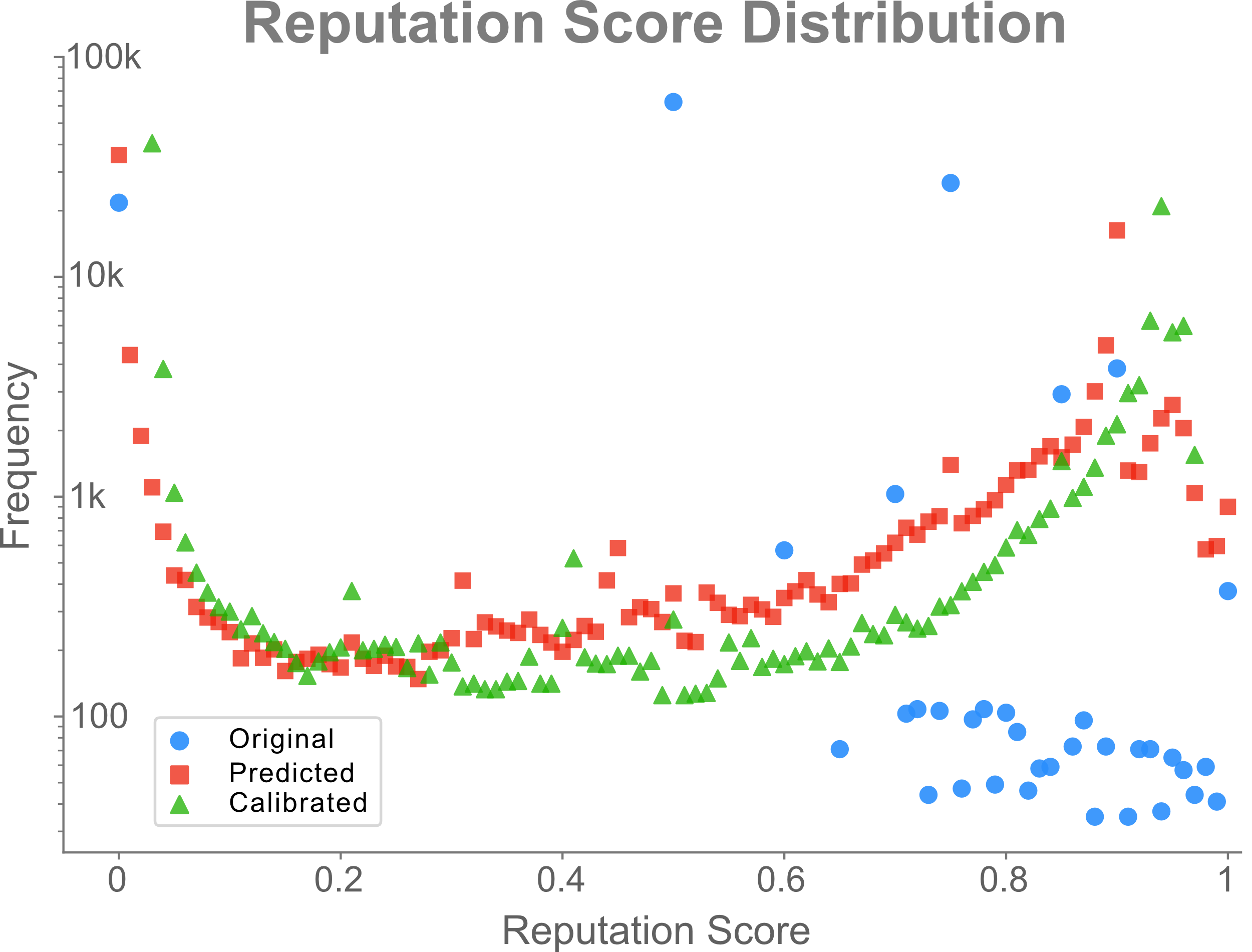}
    \caption{Reputation propagation distributes scores to neighboring entities, reducing unknowns. 
    Calibration smooths the distribution and reduces the frequency of ambiguous mid-range scores}
    \label{fig:risk-distribution}
\end{figure}

\begin{table*}[ht]
\centering
\setlength{\tabcolsep}{6.8pt}
\begin{tabular}{lrrrrrrrrrrrrrrrr} 
% \toprule
\multicolumn{10}{c}{\textbf{Graph Statistics}} & \multicolumn{7}{c}{\textbf{Performance}} \\
\cmidrule(lr){1-10} \cmidrule(lr){11-17}
\textbf{R} & $\mathbf{|V|}$ & $\mathbf{|E|}$ & $\mathbf{|O|}$ & $\mathbf{|I|}$ & $\mathbf{|A|}$ & $\mathbf{|T|}$ & $\mathbf{\% U}$ & $\mathbf{\% B}$ & $\mathbf{\% M}$ & \textbf{\# Train} & \textbf{\# Val} & \textbf{\# Test} & \textbf{Pr} & \textbf{Re} & \textbf{F1} & \textbf{AUC} \\
\midrule
1 & 5.7M & 21M & 9.9k & 285k & 590k & 4.9M & 42 & 56 & 2 & 2M & 282k & 563k & .96 & .92 & .94 & .97 \\
2 & 5.5M & 17M & 9.7k & 300k & 564k & 4.6M & 41 & 57 & 2 & 1.9M & 274k & 547k & .94 & .91 & .93 & .96 \\
3 & 5.3M & 15M & 10k & 146k & 290k & 4.9M & 42 & 56 & 3 & 2M & 283k & 565k & .95 & .93 & .94 & .97 \\
4 & 2.9M & 10M & 14k & 95k & 219k & 2.5M & 54 & 44 & 2 & 819k & 117k & 234k & .92 & .96 & .94 & .97 \\
5 & 2.5M & 7.4M & 7.6k & 141k & 282k & 2M & 41 & 56 & 3 & 841k & 120k & 240k & .95 & .84 & .89 & .92 \\
6 & 2.3M & 6.6M & 18k & 166k & 407k & 1.8M & 45 & 53 & 2 & 678k & 97k & 194k & .94 & .81 & .87 & .94 \\
7 & 879k & 4.4M & 8.6k & 54k & 140k & 677k & 36 & 57 & 7 & 305k & 44k & 87k & .94 & .96 & .95 & .98 \\
8 & 856k & 3.5M & 9.5k & 56k & 148k & 642k & 49 & 48 & 3 & 230k & 33k & 66k & .82 & .91 & .86 & .93 \\
9 & 597k & 1.4M & 2.3k & 16k & 33k & 546k & 46 & 53 & 1 & 209k & 30k & 60k & .88 & .81 & .84 & .90 \\
10 & 454k & 1.2M & 1.7k & 13k & 24k & 415k & 16 & 83 & 1 & 245k & 35k & 70k & .89 & .76 & .82 & .90 \\
11 & 38k & 120k & 398 & 1.1k & 2.1k & 35k & 35 & 63 & 2 & 16k & 2.3k & 4.5k & .82 & .84 & .83 & .87 \\
12 & 18k & 35k & 443 & 1.2k & 2.5k & 14k & 24 & 68 & 8 & 7.3k & 1.1k & 2.1k & .99 & .83 & .91 & .96 \\
\bottomrule
\end{tabular}
\caption{Comparison of performance across 12 sampled regions ($R$). Graph statistics include the number of nodes ($|V|$), edges ($|E|$), organizations ($|O|$), incidents ($|I|$), alerts ($|A|$), and entities ($|T|$), and the distribution of unknown (\% U), benign (\% B), and malicious (\% M) entities.
Performance metrics are based on a random 70-10-20 entity split across training, validation, and test sets.
Precision, recall, and F1 are reported on the test set with macro weighting, and PR-AUC with micro weighting.}
\vspace{-4mm}
\label{table:combined-results}
\end{table*}

\subsection{Reputation Scoring}\label{subsec:reputation-scoring}
To evaluate the effectiveness of \methodshort's reputation scoring, we focus on three key questions: (1) does the LP algorithm converge to produce stable and accurate reputation scores;
(2) how does scale and variation in SOC telemetry across different geographic regions affect \methodshort's predictive capability; and (3) how does \methodshort{} perform across different entity types? 
Each of these questions is addressed in the following paragraphs.

\begin{figure}[t]
    \centering
    \includegraphics[width=\linewidth]{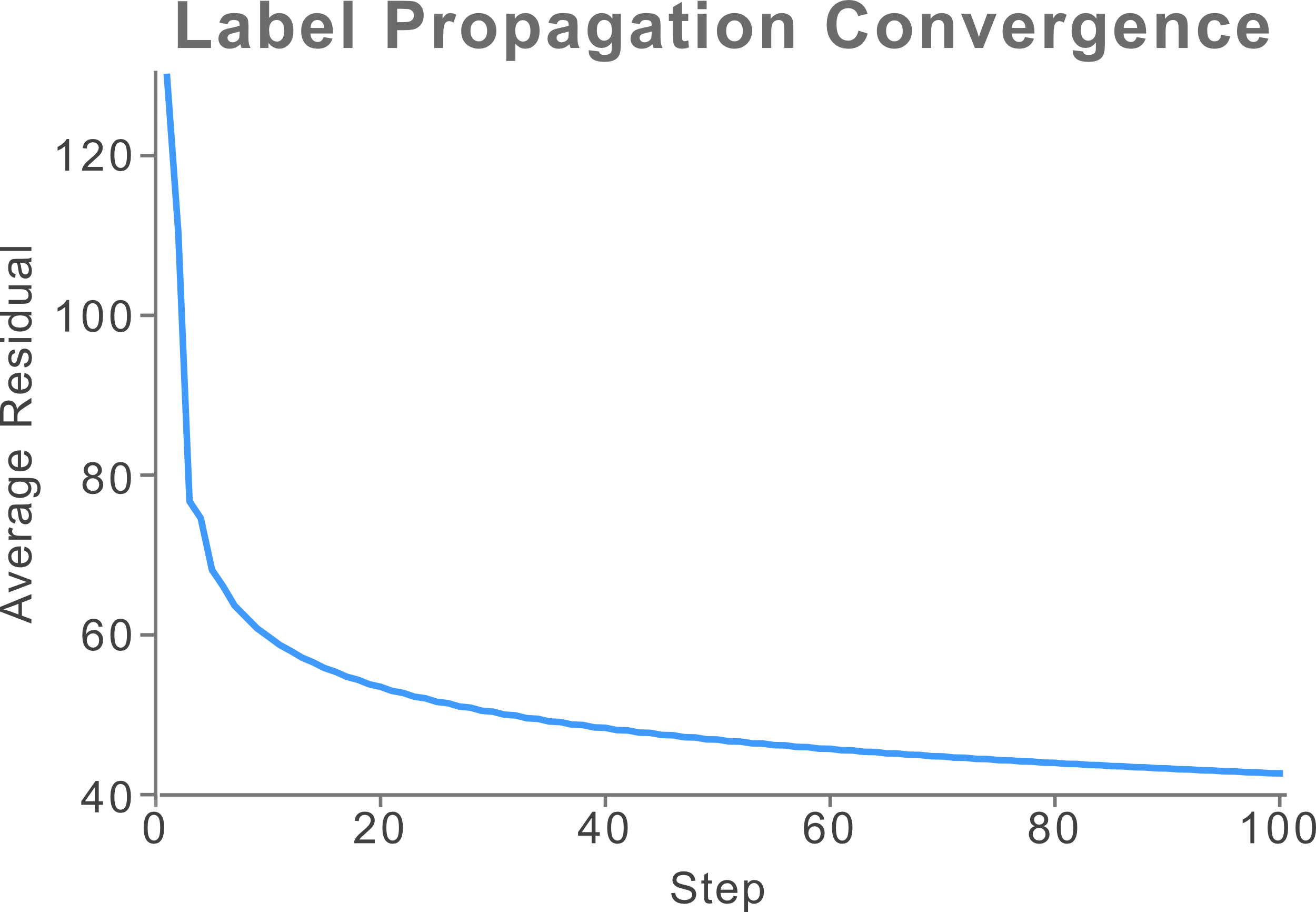}
    \caption{Convergence of label propagation measured by the average Frobenius norm residual across 466 runs.}
    \label{fig:convergence}
    \vspace{-4mm}
\end{figure}

\smallskip\noindent
\textbf{Propagation convergence.}
We analyze the convergence of the LP algorithm by tracking the average Frobenius norm residual across $466$ cross-region runs.
Figure~\ref{fig:convergence} shows a sharp initial decrease in the residual, then gradually flattens as it approaches a steady state, approaching convergence after approximately $20$ steps.
Figure~\ref{fig:risk-distribution} shows the impact of reputation propagation, where original reputation scores (blue) display a trimodal pattern concentrated in the middle and at the extremes, while propagated scores (red) exhibit a smoother distribution, reflecting reduced uncertainty.

\smallskip\noindent
\textbf{Cross-region meta analysis.}
Table~\ref{table:combined-results} presents a comprehensive overview of \methodshort's performance across a variety of geographic regions, graph sizes, and network complexities.
Each row provides a snapshot of the dynamic graph at its peak support level within the five-day window illustrated in Figure~\ref{fig:graph-size}. 
To ensure a robust analysis, our evaluation includes entities with a minimum of $100$ ground-truth reputation scores, covering IPs, email messages, email clusters, email addresses, and file hashes. 
Given the prevalence of edge types outlined in Table~\ref{tab:edge-types}, these are the entities expected to have a high volume of ground truth.

Results show that \methodshort{} consistently achieves high performance across threat landscapes, with an average macro-F1 score of $0.89$ and precision-recall AUC of $0.94$. 
These threat landscapes include: (1) large, complex graphs with millions of nodes and edges, (2) smaller, simpler graphs with thousands of nodes and edges, and (3) a range of label distributions, with $16\%$ to $54\%$ of nodes unlabeled.
In general, performance tends to decline in smaller regions, where limited contextual information can constrain effective reputation propagation, and in regions with a low proportion of malicious entities, making it challenging to identify rare positive cases.

\begin{figure}[t]
    \centering
    \includegraphics[width=1\linewidth]{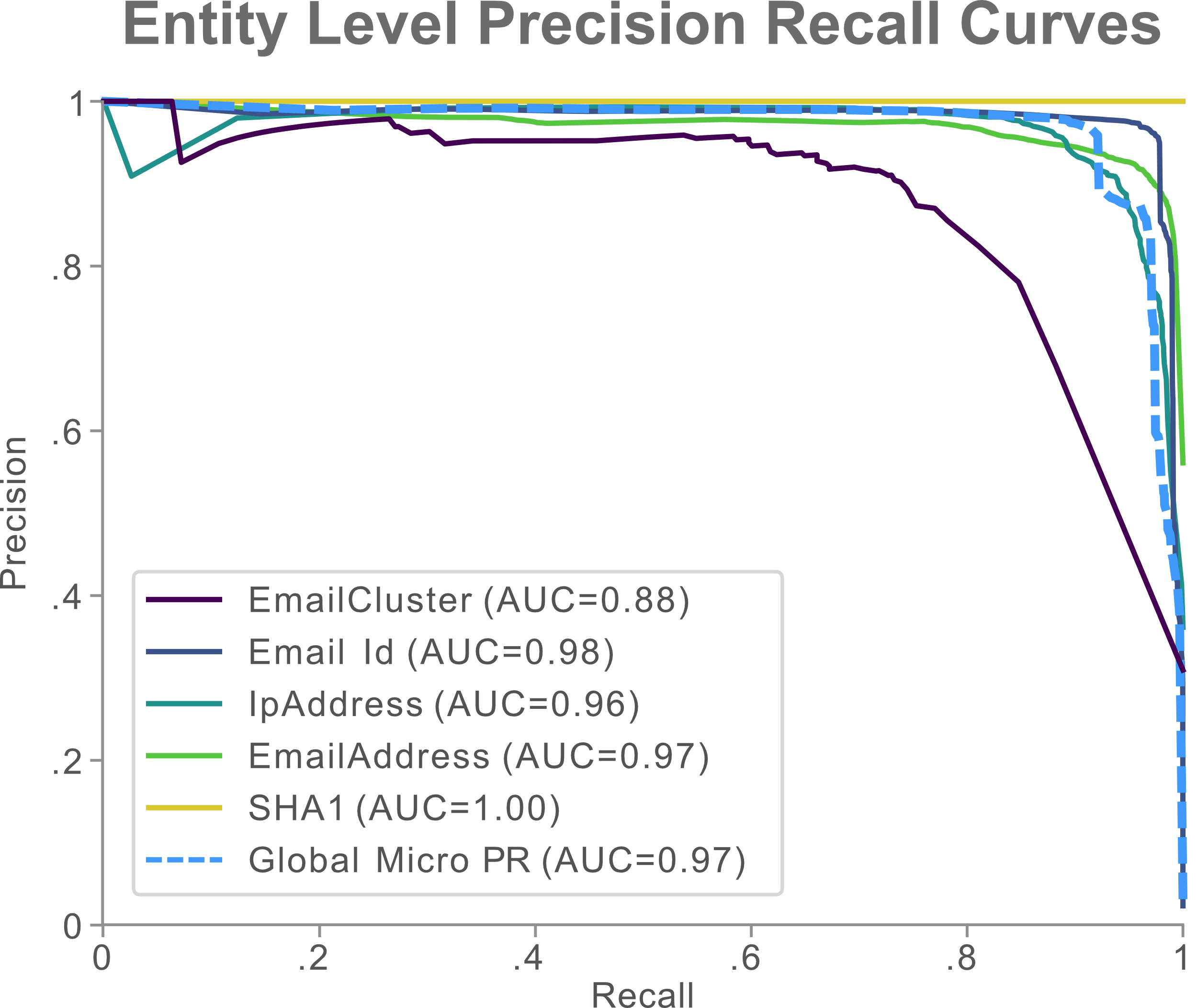}
    \caption{Precision-recall curves for each entity type in Region 1 with at least 100 ground-truth examples. 
    The global micro PR curve represents the average detection performance across all entities.}
    \label{fig:pr-curves}
    \vspace{-2mm}
\end{figure}

\begin{figure}[t]
    \centering
    \includegraphics[width=\linewidth]{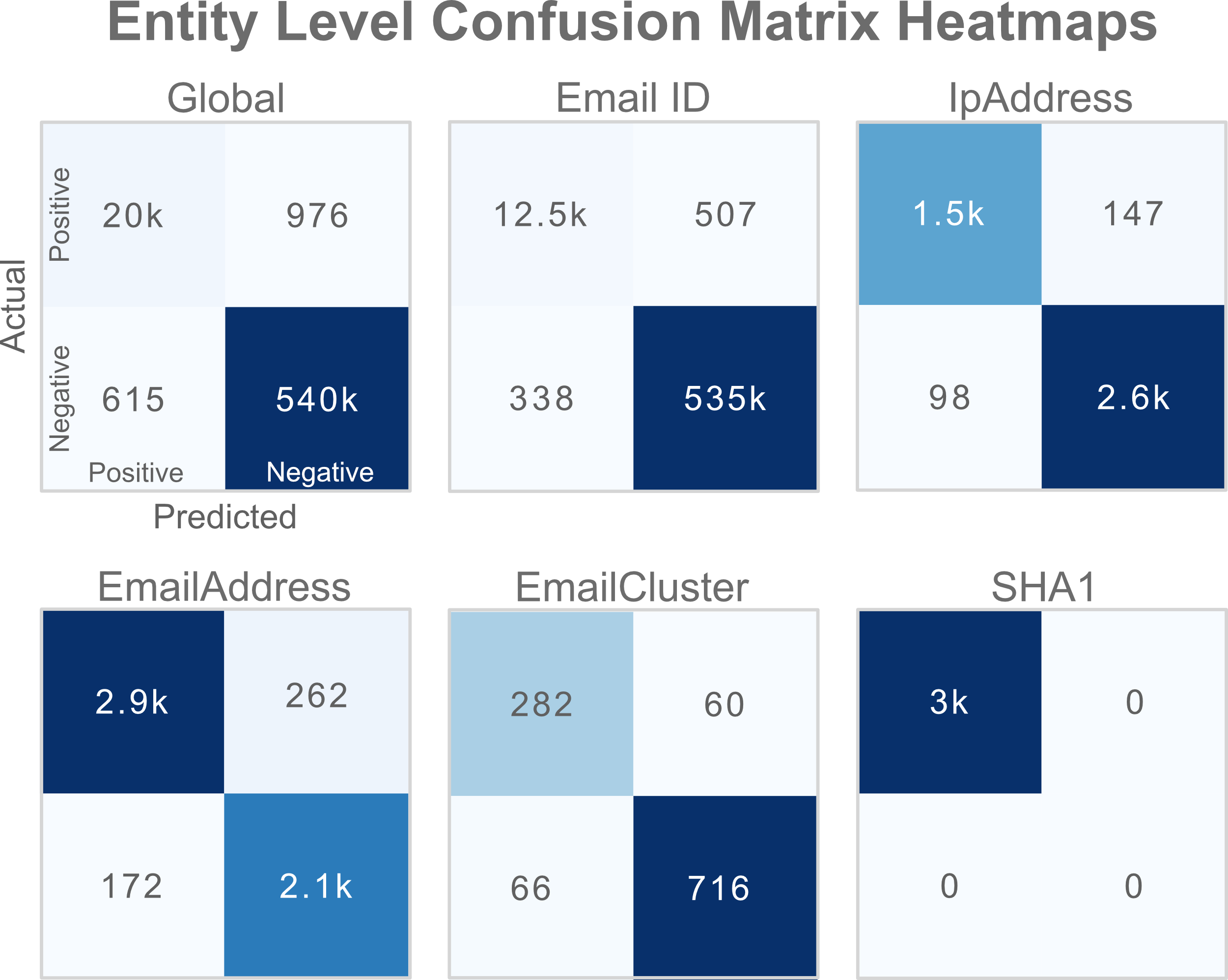}
    \caption{Confusion matrix heatmaps showing detection performance across entity types in Region 1. 
    The global heatmap represents overall performance across entity types.
    }
    \label{fig:entity-cms}
    \vspace{-4mm}
\end{figure}

\smallskip\noindent
\textbf{Per-entity granular analysis.}
To evaluate \methodshort's effectiveness at a granular level, we examine performance across individual entity types in Region 1. 
The precision-recall curves in Figure~\ref{fig:pr-curves} demonstrate that \methodshort{} consistently achieves high AUC scores across entity types, with AUCs ranging from $0.88$ to $1$. 
The AUC for EmailCluster is slightly lower than for other entities, likely due to the greater complexity and variability in the clustering of emails by the product.
The confusion matrices in Figure~\ref{fig:entity-cms}, representing the highest F1 score point on each PR curve, illustrate \methodshort’s ability to identify both malicious and benign activity across entity types.

\subsection{Model Calibration}\label{subsec:model-calibration}
The optimal temperature is calculated per region and job run on the validation set, and then applied across the graph to produce calibrated probabilities. 
Figure~\ref{fig:risk-distribution} shows the effect of calibration on the reputation score distribution, where calibrated scores (green) show a smoother distribution, reducing the frequency of ambiguous mid-range scores found in uncalibrated predictions (red). 
Across $12$ geographic regions and $466$ runs, the average temperature coefficient is $0.26$. 
Calibration achieves an average reduction of $9.54\%$ in negative log-likelihood, with score changes ranging from $4\%$ to $19\%$ across entities and regions, resulting in an average reputation score shift of $8.6\%$. 
Table~\ref{tab:calibration} provides a detailed breakdown of calibration impact across entity types and regions.

\begin{table}[t]
\centering
\setlength{\tabcolsep}{3.2pt}
% \footnotesize
\begin{tabular}{lrrrrrrrrrrrrr}
% \toprule
& \multicolumn{12}{c}{\textbf{Region}} \\
\cmidrule(lr){2-13}
\textbf{Entity} & \textbf{1} & \textbf{2} & \textbf{3} & \textbf{4} & \textbf{5} & \textbf{6} & \textbf{7} & \textbf{8} & \textbf{9} & \textbf{10} & \textbf{11} & \textbf{12} & \textbf{Avg} \\
\midrule
SHA1 & 4 & 10 & 12 & 6 &  7 & 5 & 5 & 6 & 11 & 16 & 9 & 7 & 8 \\
EmailId  & 5 & 11 & 10 & 7 & 10 & 6 & 5 & 7 & 11 & 19 & 9 & 5 & 9 \\
IpAddress & 7 & 11 & 11 & 7 &  9 & 7 & 7 & 7 & 11 & 15 & 10 & 8 & 9 \\
EmailAddress & 7 & 10 &  8 & 6 &  9 & 6 & 6 & 7 & 11 & 18 & 8 & 7 & 9  \\
EmailCluster & 3 & 10 & 10 & 5 &  8 & 4 & 3 & 5 & 11 & 18 & 10 & 6 & 8 \\

\bottomrule
\end{tabular}
\caption{Calibration of reputation scores by entity and region.
}
\vspace{-3mm}
\label{tab:calibration}
\end{table}

\section{Deployment and Impact}\label{sec:deployment}
\textbf{Deployment.} 
\methodshort{} is deployed globally, supporting hundreds of thousands of Microsoft Unified Security Operations Platform customers over the past few months. 
To comply with privacy regulations, \methodshort{} is uniformly replicated across geographic regions using Synapse.
PySpark’s distributed computational engine enables efficient large-scale data preprocessing, while Python handles graph construction and inference in areas where PySpark support is limited.
Infrastructure comprises three main components: 
(a) an ADLS database that ensures both accessibility and secure management of alert telemetry;
(b) an Azure Synapse backend  that provides a robust deployment and monitoring framework, essential for large-scale, real-time processing;
and (c) an XXL PySpark pool comprised of 60 executors, each equipped with 64 CPU cores and 400GB of RAM.
To maintain continuous and comprehensive coverage, \methodshort{} runs every hour in each region.
To enhance system reliability and prevent potential coverage gaps, Synapse reruns any failed jobs.

\smallskip\noindent
\textbf{Impact.} 
\methodshort{} processes billions of alerts each month, transforming them into actionable intelligence that drives critical detection and disruption capabilities.
Each week, \methodshort{} identifies millions of high-risk entities, enabling a 6x increase in non-file threat intelligence.
One of the key metrics in the USOP space is the incident disruption ratio, which measures the percentage of incidents that can be automatically mitigated before adversaries have a chance to cause harm. 
Leveraging \methodshort’s threat intelligence, Microsoft USOP has increased its disruption rate by $21\%$---with an average time to disrupt 1.9x faster than before---helping prevent costly breaches and mitigate risk for enterprise customers.
Additionally, \methodshort{} demonstrates remarkable accuracy, achieving 99\% precision in disrupting threats as verified by customer feedback and in-depth investigations by our threat research team.

\section{Conclusion}
\methodshort{} represents a groundbreaking advancement in enterprise cybersecurity, marking the first time a security company has openly discussed a deployed threat intelligence platform that safeguards the entire USOP landscape.
By introducing innovations such as a dynamic k-partite graph, reputation propagation mechanisms, and the integration of security domain expertise, \methodshort{} is able to uncover millions of hidden threat actor infrastructure components each week.
With an impressive average macro-F1 score of $0.89$ and a precision-recall AUC of $0.94$, \methodshort{} identifies millions of high-risk entities each week, achieving a 6x increase in non-file threat intelligence. 
Integrated into Microsoft Unified Security Operations Platform, which is deployed to hundreds of thousands of organizations globally, \methodshort{} has contributed to a $21\%$ increase in the overall USOP incident disruption rate, while reducing the time to disrupt by a factor of 1.9x. 
With a $99\%$ accuracy confirmed through customer feedback and extensive manual evaluation of thousands of incidents by security experts, \methodshort{} plays a critical role in protecting customers from costly security breaches.

\begin{acks}
We thank all of our colleagues who supported this research.
\end{acks}

\bibliographystyle{ACM-Reference-Format}
\bibliography{main.bib}

\end{document}